\pdfoutput=1
\documentclass[aps,prd,twocolumn,superscriptaddress,preprintnumbers,floatfix,nofootinbib,notitlepage,showkeys,showpacs]{revtex4-1}

\usepackage[utf8]{inputenc}
\usepackage{cancel}

\usepackage{graphicx}
\usepackage{hyperref}
\usepackage{latexsym}
\usepackage{amsmath}
\usepackage{amssymb}
\usepackage{bbm}
\usepackage{microtype}
\usepackage{nicefrac}

\usepackage[normalem]{ulem}
\usepackage{pdfsync}
\usepackage{epsfig}
\usepackage{epstopdf}
\usepackage{subfigure}
\usepackage{color}
\usepackage{comment}
\usepackage{slashed}
\usepackage{placeins}
\usepackage{cleveref}

\usepackage{xcolor}
\definecolor{cites}{RGB}{0,180,0}
\definecolor{links}{RGB}{200,0,0}
\hypersetup{colorlinks=true,citecolor=cites,linkcolor=links,urlcolor=blue}

\usepackage{graphicx}

\usepackage{tabu}
\usepackage{float}

\newcommand{\beq}{\begin{eqnarray}}
\newcommand{\eeq}{\end{eqnarray}}

\newcommand{\bmp}{\noindent\begin{minipage}{16cm}}
\newcommand{\emp}{\end{minipage}\vskip 7mm} 

    \newcommand{\ii}{\mathrm{i}}
    \newcommand{\dd}{\mathrm{d}}

    \newcommand{\bee}{\begin{equation}}
        \newcommand{\eee}{\end{equation}}

\def\lsim{\mathrel{\rlap{\lower4pt\hbox{\hskip1pt$\sim$}}
    \raise1pt\hbox{$<$}}}                
\def\gsim{\mathrel{\rlap{\lower4pt\hbox{\hskip1pt$\sim$}}
    \raise1pt\hbox{$>$}}}                

\usepackage{tikz}
\usetikzlibrary{arrows,shapes.misc,decorations.markings,decorations.pathmorphing,positioning,intersections,patterns}
\tikzset{cross/.style={cross out, draw=black, minimum size=2*(#1-\pgflinewidth), inner sep=0pt, outer sep=0pt},
cross/.default={1.5mm}}
\tikzset{mydash/.style={dashed, dash pattern=on 4pt off 5pt}}
\tikzset{
  vertex/.style={draw,shape=circle,fill=black,minimum size=3pt,inner sep=0pt},
  cross/.style={cross out, draw=black,thick, minimum size=6pt, inner sep=0pt, outer sep=0pt},
  cross2/.style={path picture={\draw[black](path picture bounding box.south east) -- (path picture bounding box.north west) (path picture bounding box.south west) -- (path picture bounding box.north east);
}}
  external/.style={inner sep=2pt},
  plabel/.style={inner sep=2pt},
  blob/.style={circle,minimum size=0.7cm,fill=black!20,draw,thick,pattern=north west lines, pattern color=black!20},
  dblob/.style={circle,size=5pt,fill=black!20,draw,thick,pattern=north west lines, pattern color=black!20},
  whiteblob/.style={circle,fill=white,minimum size=1.0cm,draw,thick},
  whiteblob2/.style={circle,fill=white,minimum size=1.0cm},
  effective/.style={rectangle,fill=black!20,minimum size=0.5cm,draw,thick},
  vev/.style={shape=vev,draw,inner sep=2pt,thick},
  mass/.style={shape=cross,draw,thick},
  rscalar/.style={dashed,thick},
  mfermion/.style={thick},
  scalar/.style={postaction={decorate}, decoration={markings,mark=at position .55 with {\arrow{latex}}},dashed,thick},
  ooscalar/.style={postaction={decorate}, decoration={markings,mark=at position .7 with {\arrow{latex}}},dashed,thick},
  fermion/.style={postaction={decorate}, decoration={markings,mark=at position .55 with {\arrow{latex}}},thick},
  majfermion/.style={postaction={decorate}, decoration={markings,mark=at position .7 with {\arrow{latex}}},thick},
  oofermion/.style={postaction={decorate}, decoration={markings,mark=at position .85 with {\arrow{latex}}, mark=at position .35 with {\arrowreversed{latex}}},thick},
  iifermion/.style={postaction={decorate}, decoration={markings,mark=at position .35 with {\arrowreversed{latex}}, mark=at position .85 with {\arrow{latex}}},thick},
  gaugeboson/.style={decorate, decoration={snake},thick},
  gluon/.style={decorate, decoration={coil,amplitude=4pt, segment length=5pt},thick},
  photon/.style={decorate, decoration={snake},thick},
  dashdot/.style={dash pattern=on .4pt off 3pt on 4pt off 3pt,thick}
}

\begin{document}
\title{
Bubble-resummation and critical-point methods for $\beta$-functions at large $N$ }
\author{Tommi Alanne}
\email{tommi.alanne@mpi-hd.mpg.de}
\affiliation{Max-Planck-Institut f\"{u}r Kernphysik, 
    Saupfercheckweg 1, 69117 Heidelberg, Germany}
\author{Simone Blasi}
\email{simone.blasi@mpi-hd.mpg.de}
\affiliation{Max-Planck-Institut f\"{u}r Kernphysik, 
    Saupfercheckweg 1, 69117 Heidelberg, Germany}
\author{Nicola Andrea Dondi}
\email{dondi@cp3.sdu.dk}
\affiliation{CP$^3$-Origins, University of Southern Denmark, Campusvej 55, 5230 Odense M, Denmark}

\begin{abstract}
We investigate the connection between the bubble-resummation and critical-point 
methods for computing the $\beta$-functions in the limit of large number of flavours, $N$, 
and show that these can provide complementary information.
While the methods are equivalent for single-coupling theories, 
for multi-coupling case the standard critical 
exponents are only sensitive to a combination of the independent
pieces entering the $\beta$-functions, so that additional input or direct computation are needed to 
decipher this missing information. 
In particular, we evaluate the $\beta$-function for the quartic 
coupling in the Gross--Neveu--Yukawa model, thereby completing 
the full system at $\mathcal{O}(1/N)$.
The corresponding critical exponents would imply a shrinking radius of convergence when $\mathcal{O}(1/N^2)$ 
terms are included, but our present result shows that the new 
singularity is actually present already at $\mathcal{O}(1/N)$, when the full 
system of $\beta$-functions is known.
\end{abstract}


\preprint{CP3-Origins-2019-16 DNRF90}

\maketitle
\section{Introduction}

The computation of the RG functions in the limit of large number of flavours, $N$, has been traditionally divided into two schools: 
{\it (i)}~the direct computation of the $\beta$-functions in a fixed space-time dimension resumming specific classes 
of diagrams in the perturbative expansion around the Gau\ss ian fixed point
~\cite{Espriu:1982pb,PalanquesMestre:1983zy,Kowalska:2017pkt,Antipin:2018zdg,Alanne:2018ene,Alanne:2018csn}, 
and {\it (ii)}~evaluation of the critical exponents at 
the Wilson--Fisher fixed point in $d$ dimensions of theories in the same universality class, 
see e.g. Refs~\cite{Vasiliev:1981yc,Vasiliev:1981dg,Vasiliev:1982dc,Gracey:1993ua,Gracey:1996he,Ciuchini:1999wy,Gracey:1990wi,Gracey:1992cp,Derkachov:1993uw,Vasiliev:1992wr,Vasiliev:1993pi,Gracey:1993kb,Gracey:1993kc,Gracey:2017fzu,Manashov:2017rrx}
and Ref.~\cite{Gracey:2018ame} for a recent review. 

In particular for one-coupling systems, the critical-point formalism is very powerful 
since in this case the $\beta$-function can be computed once its slope at criticality is known; 
these results were recently also used to assess the 
apparent singularity structure of gauge $\beta$-functions~\cite{Ryttov:2019aux}. 
Furthermore, the method is technically more convenient beyond the leading order in $1/N$, 
even though attempts to reconstruct the leading singularity through high-order analysis are 
ongoing~\cite{Dondi:2019ivp}.

Conversely, the bubble-resummation method is more versatile: As we will show, for multi-coupling systems, the knowledge of the
critical exponents is not enough to reconstruct the full system of $\beta$-functions, and one needs to
either input 
additional information or rely on 
a direct computation. Furthermore, the bubble-resummation method has recently been used to
compute other quantities beyond the various RG functions, 
like conformal anomaly coefficients at large-$N$~\cite{Antipin:2018brm}.

The purpose of this paper is to compare these two methods and to show that they can provide 
complementary information. We will first consider a generic one-coupling system, 
and provide a dictionary between these two methods; see e.g. 
Ref.~\cite{Ferreira:1997he} for a similar attempt in the context of Wess--Zumino model,
Ref.~\cite{Friess:2005be} for 2D non-linear sigma models in a string theory context
and Ref.~\cite{Gracey:2018ame} for the general $\mathcal{O}(1/N)$ result.
As a prime example, we will consider the Gross--Neveu (GN) model, 
whose critical exponents have been extensively studied; 
see e.g. Refs~\cite{Gracey:1990wi,Vasiliev:1992wr,Gracey:1993kb,Gracey:1993kc}.  
In particular, the slope of the $\beta$-function is known at $\mathcal{O}(1/N^2)$.
We will extract the explicit $\beta$-function up to this order and comment on the possibility of an 
IR fixed point for two-dimensional GN model~\cite{Schonfeld:1975us,Choi:2016sxt},
which turns out to be disfavoured.

Secondly, we will explicitly show the complementarity of the two methods
in the context of a two-coupling system, namely
the Gross--Neveu--Yukawa (GNY) model, 
where we were able to compute the full coupled system 
of $\beta$-functions at $\mathcal{O}(1/N)$ completing the 
results of Ref.~\cite{Alanne:2018ene}. 
For the GNY model, the critical exponents are also known up to
$\mathcal{O}(1/N^2)$~\cite{Gracey:2017fzu,Manashov:2017rrx}. 
We will use these as an input to derive consistency conditions for the $\beta$-functions
and gain information regarding the location of the poles at 
different orders in the expansion. Although the $\mathcal{O}(1/N^2)$ result implies an 
appearance of a new singularity with respect to the $\mathcal{O}(1/N)$ 
critical exponents, we will show that this apparent new
singularity is actually 
already present at $\mathcal{O}(1/N)$ once the full system $\beta$-functions is known, and the 
disappearance in the critical exponent is due to a subtle cancellation of different contributions.

The paper is organised as follows: In Sec.~\ref{sec:1coupling} we review the connection of the two methods in the one-coupling system and 
study the GN model in two dimensions as a concrete example. In Sec.~\ref{sec:GNY} we compute the full coupled system of $\beta$-functions for GNY model up to $\mathcal{O}(1/N)$ and relate 
our result to the known critical exponents. 
In addition, we provide the $\mathcal{O}(1/N)$ contributions
to the perturbative $\beta$-functions up to six-loop order.
In Sec.~\ref{sec:conclusions}
we provide our conclusions. Finally, in Appendix~\ref{app:N2} we give the corresponding relation between the $\beta$-functions and the critical exponents in the GNY model at $\mathcal{O}(1/N^2)$.

\section{One-coupling model}
\label{sec:1coupling}
In this section, we discuss the general ansatz for the $\beta$-function 
in the large-$N$ expansion for any system with one coupling,
$g$. Our goal is to derive a general form for the $\beta$-function
once the critical exponent $\omega = \beta^\prime(g_c)$, where $g_c$ is 
the coupling at the Wilson--Fisher fixed point, is known.
We define\footnote{In the literature, there is often an extra factor of $-2$ on the 
left-hand side of the definition, Eq.~\eqref{eq:betaprime}. We omit that here for the sake of simplicity.}
\begin{equation}
    \label{eq:betaprime}
    \beta'(g_c) = \omega(d)\equiv\sum_{n=0}^{\infty}\frac{\omega_n(d)}{N^n}\,,
\end{equation}
while the ansatz for the $\beta$-function is:
\begin{equation}
    \label{eq:1couplbeta}
    \beta(g) = (d-d_c)g + g^2 \left( b N + c + 
    \sum_{n=1}^\infty \frac{F_n(g N)}{N^{n-1}}\right),
\end{equation}
where $d$ is the dimension of the space-time, $d_c$ the critical dimension of the coupling $g$, 
$b$ and $c$ are model-dependent one-loop coefficients, and $F_n$ are resummed functions satisfying $F_n(0) = 0$.

Requiring $\beta(g_c) = 0$, we find an implicit expression for the 
critical coupling:
\begin{equation}\label{eq:gc}
     g_c = -\frac{d-d_c}{b N + c+ \sum_{n=1}^\infty \frac{F_n(g_c N)}{N^{n-1}}}.
\end{equation}
The slope of the $\beta$-function at criticality 
can then be expanded in $1/N$ to yield
\begin{equation}\label{eq:1cslope}
    \begin{split}
	&\beta^\prime(g_c) = -(d-d_c) +  \frac{(d-d_c)^2}{b^2}
	    \sum_{m=1}^\infty \frac{F^\prime_m(g_c N)}{N^m} \\
	&\quad \times \sum_{k=0}^\infty (-b)^{-k} (k+1)\left( \frac{c}{N} + \sum_{n=1}^\infty
	    \frac{F_n(g_c N)}{ N^n} \right)^k.
    \end{split}
\end{equation}
Using Eqs~\eqref{eq:betaprime} and \eqref{eq:1cslope}, we can relate the functions
$F_n$ to $\omega_n$.
To obtain the result in a closed form, it is necessary to
compute $g_c$ order by order in $1/N$ according to
Eq.~\eqref{eq:gc}. This, in turn, enters the argument of the functions $F_n$, which then
need to be Taylor-expanded to include all
the relevant contributions.
In the following, we will give 
explicitly the first two orders. 
At $\mathcal{O}(1/N)$, we obtain
\begin{equation}
    \omega_1(d) = \beta^\prime_1(g_c) = \frac{ (d-d_c)^2}{b^2}
    F_1^\prime \left(\frac{d_c-d}{b}\right),
\end{equation}
which, defining $t\equiv(d_c-d)/b$, results in 
\begin{equation}
    F_1(K) = \int_0^K \dd t \frac{\omega_1\left(d_c-bt\right)}{t^2}.
\end{equation}
At $\mathcal{O}(1/N^2)$, the expansion of Eq.~\eqref{eq:1cslope} gives
\begin{equation}
\label{eq:F2_1coupl}
\begin{split}
    F_2(K) = \int_0^K \dd t
    &\left(
     \frac{c + F_1(t)}{b}(t F_1^{\prime\prime}(t) + 2 F_1^\prime(t))\right.\\
    &\left.\quad+  \frac{\omega_2\left(d_c-bt\right)}{t^2} \right).
\end{split}
\end{equation}
Note that the critical exponent $\omega_1$ contributes to 
the $\beta$-function also beyond $\mathcal{O}(1/N)$ through $F_1$ and its derivatives,
as can be seen explicitly in Eq.~\eqref{eq:F2_1coupl}.
The same structure is found at
higher orders: $F_n$ receives contributions
from $\omega_{n-1},\dots,\omega_1$---or, equivalently,
from $F_{n-1},\dots,F_1$---and their derivatives,
together with a pure $\omega_n$-term
as in the last line of Eq.~\eqref{eq:F2_1coupl}.
Therefore, if $F_1$ has a singularity say at 
$K = K_s$, it will propagate to $F_n$
with a stronger degree of divergence up to
the $n$-th derivative of $F_1$. 
This confirms the expectation that the singular
structure of the higher-order $F_n$ functions
contain all the
singularities of the lower ones, together
with new possible singularities brought in
by the pure $\omega_n$-contribution.
On the other hand, the fact that the illustrative
singularity at $K = K_s$ would appear at any order in the $1/N$-expansion suggests
that a resummation could exist such that the $\beta$-function is regular at $K_s$.

\subsection{Gross--Neveu model in $d=2+\epsilon$}
The possibility of an IR fixed point in the 
GN model in $2+\epsilon$ dimensions was recently studied~\cite{Choi:2016sxt} 
using the perturbative four-loop result~\cite{Gracey:2016mio} with Pad\'{e} approximants. 
On the other hand, the presence of an IR fixed point 
in the large-$N$ limit has already been excluded taking the $\mathcal{O}(1/N)$ contributions into 
account~\cite{Schonfeld:1975us}. In this section, we extend the analysis to
$\mathcal{O}(1/N^2)$ by using the results of the previous section and the 
known results for the critical exponent, $\lambda(d)$,
\begin{equation}
    \label{eq:}
     \lambda(2+\epsilon)\equiv\sum_{n=0}^{\infty}\frac{\lambda_n(\epsilon)}{N^n} = \beta^{\prime}(g_c),
\end{equation}
which is currently known up to $\mathcal{O}(1/N^2)$~\cite{Gracey:1993kb}.
The $\mathcal{O}(1/N)$ coefficient is explicitly given by
\begin{equation}
    \label{eq:}
    \lambda_1(t)=-\frac{2t\Gamma(t+2)\sin(\pi t/2)}{\pi(t+2)\Gamma(t/2+1)^2},
\end{equation}
while the expression for $\lambda_2(t)$ 
is relatively lengthy and can be explicitly found in Ref.~\cite{Gracey:1993kb}. 

Referring to Eq.~\eqref{eq:1couplbeta}, the GN model is
characterized by $d_c = 2$, $b=-1$ and $c = 2$.
However, we modify the ansatz of 
Eq.~\eqref{eq:1couplbeta} to implement the fact that for $N=2$ the GN model is equivalent to the 
abelian Thirring model~\cite{Thirring:1958in}, and thus the $\beta$-function identically 
vanishes~\cite{Mueller:1972md,Gomes:1972yb}:
\begin{equation}
    \label{eq:ansatz2}
    \begin{split}
	&\beta(g) = (d-2)g \\
	&\ \ + (N-2)g^2 \left( -1 + \frac{\tilde{F}_1(g N)}{N}+ \frac{\tilde{F}_2(g N)}{N^2} + \dots \right),
    \end{split}
\end{equation}
where 
\begin{equation}\label{eq:F1GN}
    \tilde{F}_1(K) = -2\int_0^K \frac{\lambda_1(t)}{t^2} \dd t
\end{equation}
and
\begin{equation}\label{eq:F2GN}
    \begin{split}
	\tilde{F}_2(K) =  \int_0^K &\left\{\frac{ - 2 \lambda_2(t) + 4 \lambda_1(t) +
	4 \lambda_1(t) F_1(t)}{t^2}\right.\\
	&\left.\vphantom{\frac12}\quad- t[2 + F_1(t)] F_1^{\prime \prime}(t) \right\}\dd t.
    \end{split}
\end{equation}
The functions $\tilde{F}_{1,2}$ are related to $F_{1,2}$ of the standard
ansatz~\eqref{eq:1couplbeta} as
\begin{equation}
    \tilde{F}_1 = F_1, \quad \tilde{F}_2 = F_2 + 2 F_1,
\end{equation}
so that the two ans\"atze coincide at $\mathcal{O}(1/N^2)$.

On the other hand, the $\beta$-function for the GN model is known perturbatively 
up to four-loop level~\cite{Gracey:2016mio}:
\begin{equation}
    \begin{split}
	\beta_{4 \mathrm{L}}(g) = & (d-2) g -(N-2) g^2  + ( N - 2)g^3\\
	&+ \frac{1}{4}(N-2)(N-7)g^4  \\
	& -\frac{1}{12}(N-2)\left[N^2+(66\zeta_3+19)N\right.\\
	&\left.\qquad\qquad\qquad- 204\zeta_3-48  \right]g^5.
    \end{split}
\end{equation}
We find that the improved ansatz, Eq.~\eqref{eq:ansatz2},
additionally reproduces the first subleading $1/N^3$ terms, in particular providing the
correct three-loop coefficient. Explicitly\footnote{Notice that while the leading $N$ coefficient is scheme 
independent~\cite{Shrock:2013cca}, the subleading ones are not. The result obtained with the critical exponent method should be 
compared with perturbation theory where $\overline{\mathrm{MS}}$ dimensional 
regularisation is employed.},

\begin{equation}
    \beta(g) - \beta_{4 \mathrm{L}}(g) =-g^5 (N-2) ( 4 + 17 \zeta(3) ) +\mathcal{O}(g^6).
\end{equation}
Furthermore, the prediction for the leading orders in $N$ based
on Eqs~\eqref{eq:F1GN} and \eqref{eq:F2GN} for the five-loop $\beta$-function 
is
\begin{equation}
    \label{eq:}
    \begin{split}
	\beta^{(5)}(g)=&\frac{1}{96}(N-2)\left[(3-6\zeta_3)N^3\right.\\
	&\left.\quad +(297\zeta_4+120\zeta_3+1)N^2+\dots\right]g^6.
    \end{split}
\end{equation}

\begin{figure}[t]
    \centering
    \includegraphics[width=0.48\textwidth]{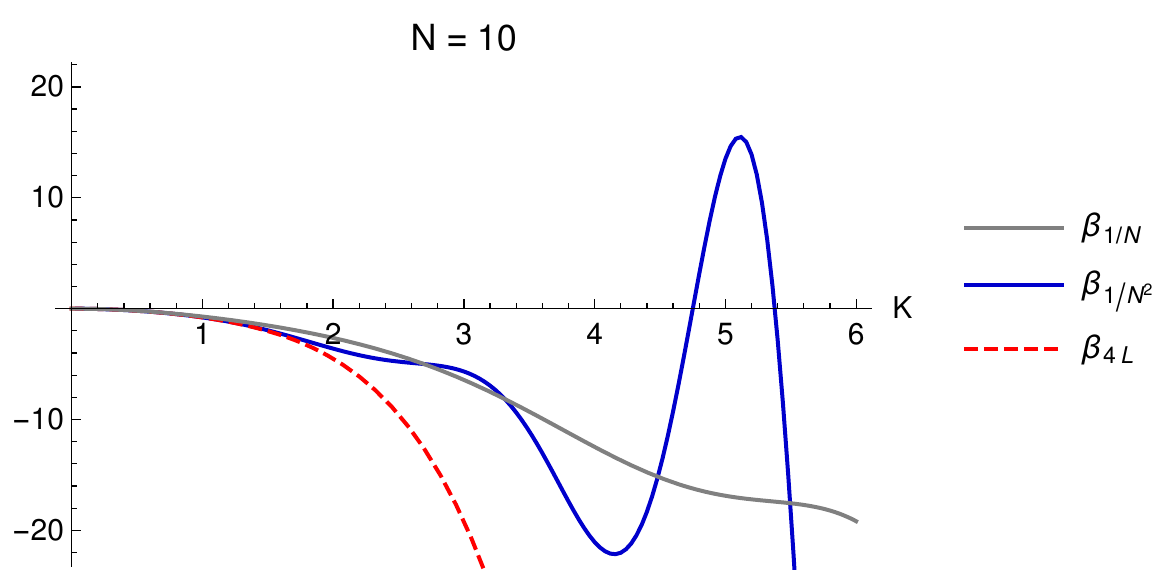}\\
    \vspace{0.5cm}
    \includegraphics[width=0.48\textwidth]{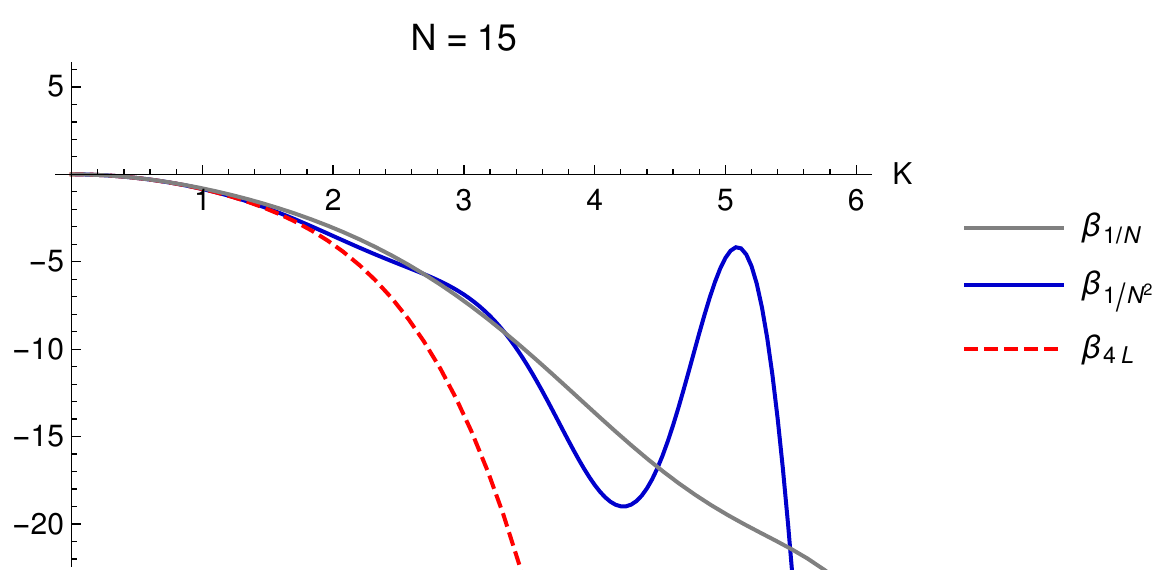}
    \caption{The  $\beta$-function of Eq.~\eqref{eq:ansatz2} truncated to $\mathcal{O}(1/N)$, $\beta_{1/N}$, 
	and to $\mathcal{O}(1/N^2)$, $\beta_{1/N^2}$ along with the four-loop perturbative result, $\beta_{4 \mathrm{L}}$, 
	as a function of the rescaled coupling $K = g N$ for $N=10,15$.}
    \label{fig:1}
\end{figure}

We show the  $\beta$-function of Eq.~\eqref{eq:ansatz2} truncated to 
$\mathcal{O}(1/N)$, $\beta_{1/N}$, and to $\mathcal{O}(1/N^2)$, $\beta_{1/N^2}$, 
along with the four-loop perturbative result
in Fig.~\ref{fig:1} as a function of the rescaled coupling $K = g N$ for $N=10,15$.
We conclude that there is no clear hint for the IR fixed point in the region where
the perturbative series is under control. 

To conclude the section, let us comment on 
the radius of convergence of the GN $\beta$-function at large $N$. The $\beta$-function does not have any singularities 
for positive couplings, although the resummed functions get contributions from graphs that grow
polynomially with the loop order. 
Therefore, one expects to find a finite radius of convergence, similarly as in e.g. QED~\cite{PalanquesMestre:1983zy}. 
In the present case, the singularities do appear, but at negative coupling values
so that the radius of convergence in the complex plane is indeed finite; this 
is related to the fact that the Wilson--Fisher 
fixed point exists above the critical dimension.
For positive coupling values this translates to a regular, though wildly oscillatory, behaviour.

\FloatBarrier

\section{Two-coupling case: Gross--Neveu--Yukawa model}
\label{sec:GNY}

\subsection{Setup}

The GNY model is the bosonised GN model with the scalar promoted to dynamical degree of freedom. It describes
$N$ massless fermion flavours, $\psi$, coupling to a massless real scalar, $\phi$, via Yukawa interaction 
\begin{equation}
    \label{eq:GNYLag}
    \mathcal{L}_{\mathrm{GNY}}=\bar{\psi}\ii \slashed{\partial}\psi-\frac{1}{2}\partial_{\mu}\phi\partial^{\mu}\phi
	+g_1 \phi\bar{\psi}\psi+g_2\phi^4.
\end{equation}
The critical dimension of the Yukawa interaction is ${d_c=4}$, and we will work in ${d=4-\epsilon}$ dimensions and 
$\overline{\mathrm{MS}}$ renormalisation scheme.
We follow the notations of Refs~\cite{Mihaila:2017ble,Zerf:2017zqi} in order to provide for a straight-forward comparison 
with the perturbative result and define rescaled couplings\footnote{We add an extra factor
of $2$ in the definition of $K$ to agree with Ref.~\cite{Alanne:2018ene}.}
\begin{equation}
    \label{eq:GNYcoupl}
    y\equiv\frac{g_1^2\mu^\epsilon}{8\pi^2},\quad K\equiv 2yN,\quad \text{and}\quad \lambda\equiv\frac{g_2\mu^\epsilon}{8\pi^2}.
\end{equation}

\subsection{The $\beta$-functions from the critical exponents}

The critical exponents, $\omega_{\pm}$, for the GNY model were recently computed up to 
$1/N^2$~\cite{Gracey:2017fzu,Manashov:2017rrx}, and on the other hand, they are known perturbatively 
up to four-loop level~\cite{Mihaila:2017ble,Zerf:2017zqi}.
The computation for the Yukawa $\beta$-function using bubble-resummation method was carried out up to $\mathcal{O}(1/N)$ in 
Ref.~\cite{Alanne:2018ene}.

The Yukawa $\beta$-function at $\mathcal{O}(1/N)$
depends only on the Yukawa coupling, $y$:
\begin{equation}\label{eq:beta1}
    \beta_y = (d-d_c) y + y^2 (2 N + 3 + F_1( y N)). 
\end{equation}
Conversely, the $\beta$-function for the quartic coupling,
$\lambda$, at $\mathcal{O}(1/N)$ is
\begin{equation}\label{eq:beta2}
    \begin{split}
	\beta_\lambda =  & (d-d_c) \lambda + y^2(-N + F_2(y N)) \\
	& + \lambda^2(36 + F_3(y N)) + y \lambda (4 N + F_4(y N)).
    \end{split}
\end{equation}
According to Eqs~\eqref{eq:beta1} and~\eqref{eq:beta2},
the coupled system of $\beta$-functions at $\mathcal{O}(1/N)$
contains four unknown functions, namely $F_1$, $F_2$, $F_3$ and $F_4$.
Note that $F_{1-4}$ are functions of the rescaled Yukawa coupling only due to the 
$1/N$ counting. Diagrammatically this corresponds to chain of fermion bubbles. Similar diagrams of scalar bubbles 
lack the $N$ enhancement, and these chains are subleading.

We can constrain $F_{1-4}$ exploiting the knowledge
of the critical exponents, $\omega_{\pm}$, by 
first determining the critical couplings such that
$\beta_{y,\lambda}=0$.
From the first equation, using $d-d_c=-\epsilon$, we find
\begin{equation}
    y^c = \frac{\epsilon}{2 N + 3 + F_1(y^c N)},
\end{equation}
and from the second
\begin{equation}
    \begin{split}
	\lambda^c =\ \frac{   \epsilon - y^c\left(4 N + F_4(y^c N)\right) + \sqrt{\Delta^c}}
	{2\left(36 + F_3(y^c N)\right)}
    \end{split}
\end{equation}
where we have taken the positive solution for $\lambda^c$ and defined
\begin{equation}
    \begin{split}
	\Delta^c \equiv &\left[- \epsilon + y^c(4 N + F_4(y^c N))\right]^2 \\
	&- 4 (36 + F_3(y^c N))(y^c)^2(-N + F_2(y^c N)).
    \end{split}
\end{equation}
Up to leading order in $1/N$, we have
\begin{equation}
 y^c= \frac{\epsilon}{2N}+\mathcal{O}(1/N^2),\quad\lambda^c = 
          \frac{\epsilon}{4 N}+\mathcal{O}(1/N^2).
\end{equation}
Since $\frac{ \partial \beta_y}{\partial \lambda} \equiv 0$, the eigenvalues
of the Jacobian, $\omega_\mp$, directly correspond to 
$\frac{ \partial \beta_y}{\partial y}$
and $\frac{ \partial \beta_\lambda}{\partial \lambda}$
at criticality, respectively.
Explicitly,
\begin{equation}\label{eq:F1}
    \frac{\partial \beta_y}{\partial y} = \epsilon  + \frac{1}{4 N} \epsilon^2 F_1^\prime(\epsilon/2)
    = \epsilon + \frac{1}{N} \omega^{(1)}_-(4-\epsilon),
\end{equation}
and
\begin{align}
	\frac{\partial \beta_\lambda}{\partial \lambda}
	=& \epsilon + \frac{\epsilon}{2 N}\left(30 - 2 F_1(\epsilon/2)
	    + F_3(\epsilon/2) + F_4(\epsilon/2)\right)\nonumber\\
	=& \epsilon + \frac{1}{N}\tilde{\omega}^{(1)}_+(\epsilon).
	\label{eq:omegap}
\end{align}
For simplicity, we denote $\tilde{\omega}_{\pm}^{(1)}(\epsilon)\equiv\omega_{\pm}^{(1)}(4-\epsilon)$ 
in the following. Equation~\eqref{eq:F1} yields
\begin{equation}\label{eq:F1GNY}
    F_1(t) = \int_0^t \frac{\tilde{\omega}^{(1)}_-(2 \epsilon)}{\epsilon^2} \text{d}\epsilon,
\end{equation}
whereas Eq.~\eqref{eq:omegap} gives
\begin{equation}\label{eq:omplus}
    30 - 2 F_1(\epsilon/2)
    + F_3(\epsilon/2) + F_4(\epsilon/2)=  2 \frac{\tilde{\omega}^{(1)}_+(\epsilon)}{\epsilon}.
\end{equation}
As we can see from Eq.\,\eqref{eq:omplus},
$\beta_\lambda$ cannot be computed
with the knowledge of $\omega_\pm$, since
only the combination $F_3 + F_4$ can be accessed. 
In particular, $F_2$ is fully unconstrained. 
This shows that the critical exponents encoding the
slope of the $\beta$-function can fully determine
the $\beta$-function only for single-coupling theory, while for
multi-coupling theory they are sensitive only to certain
combinations.
Therefore, either more information is input or
one needs to rely on a direct computation to get
$\beta_\lambda$ in a closed form.

\begin{figure}
    \begin{center}
	\includegraphics[width=0.48\textwidth]{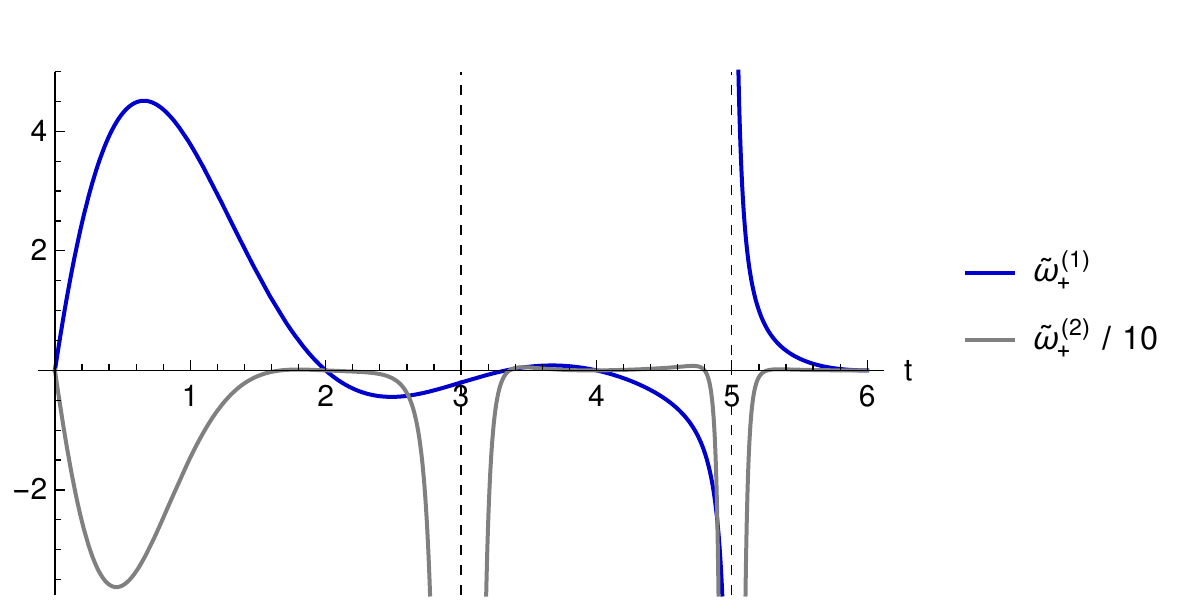}\\
	\includegraphics[width=0.48\textwidth]{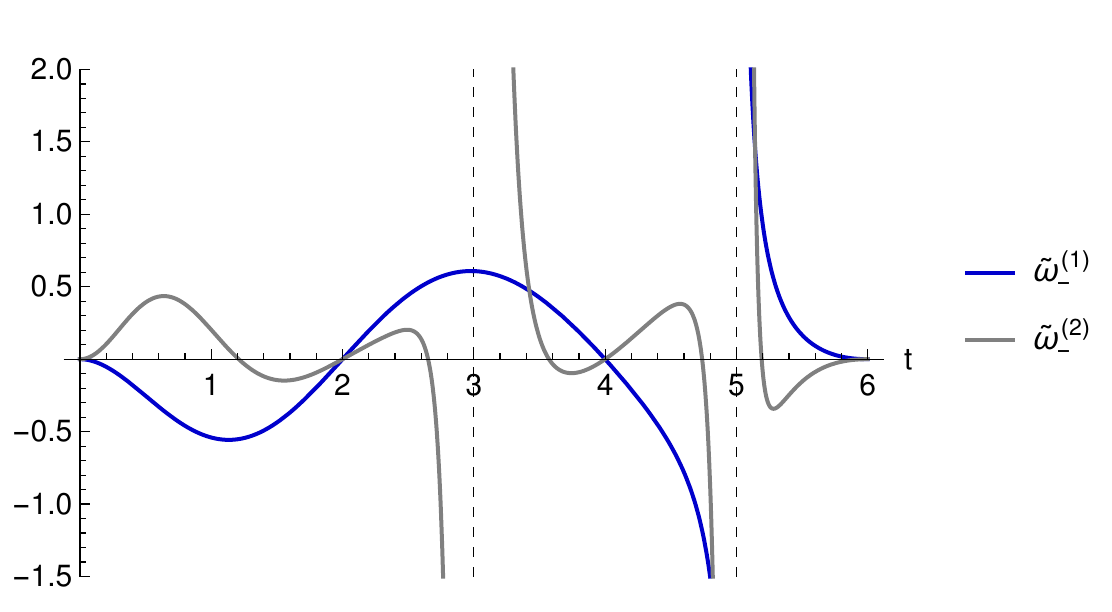}
    \end{center}
    \caption{The first two coefficients of the $1/N$ expansion of the 
	critical exponents $\tilde{\omega}_{\pm}=\sum_{n=1}^{\infty}\tilde{\omega}_{\pm}^{(n)}/N^n$. The explicit formulae for
	the $\mathcal{O}(1/N^2)$ 
	coeffiecients can be found in Ref.~\cite{Manashov:2017rrx}. 
}
    \label{fig:omegapm}
\end{figure}

Nevertheless, the knowledge
of $\omega_\pm$ can be used to obtain independent cross-checks
and gain information regarding the radius of convergence
of the $1/N$ expansion.
The explicit formulae for $\tilde{\omega}^{(1)}_\mp$ are~\cite{Gracey:2017fzu,Manashov:2017rrx}
\begin{equation}\begin{split}
 \tilde{\omega}^{(1)}_-(t) & = - \frac{t \Gamma(4-t)}
 {\pi \Gamma \left( 2 - \frac{t}{2} \right)
 \Gamma \left( 3 - \frac{t}{2} \right)} \text{sin}\left( \frac{\pi t}{2}\right), \\
 \vphantom{\frac{\frac{\frac12}{\frac12}}{\frac12}}\tilde{\omega}^{(1)}_+(t) & = \frac{3 t - 10}{t} \tilde{\omega}^{(1)}_-(t).
 \end{split}
\end{equation}
We show the critical exponents $\tilde{\omega}_{\pm}^{(1)}(t)$ 
along with the $\mathcal{O}(1/N^2)$ results~\cite{Manashov:2017rrx},
$\tilde{\omega}_{\pm}^{(2)}(t)$, in Fig.~\ref{fig:omegapm}. 
The $\mathcal{O}(1/N^2)$ results indicate 
that there is a new singularity not present at $\mathcal{O}(1/N)$ occurring 
at $t=3$. Correspondingly, this would suggest a shrinking in the radius
of convergence for the $\beta$-functions when higher orders are included.
However, as we will show in the next sections, this singularity is actually already present 
at $\mathcal{O}(1/N)$, namely in the functions $F_2,F_3,F_4$, but 
is exactly cancelled in $\tilde{\omega}_+^{(1)}$, Eq.~\eqref{eq:omplus}.  

The connection between $\beta_{y,\lambda}$ and $\omega_\mp$
at $\mathcal{O}(1/N^2)$ is enstablished in Appendix~\ref{app:N2}.
This allows us to derive conditions analogous to Eq.~\eqref{eq:omplus}
for the new unknowns parametrizing the $\beta$-functions
at $\mathcal{O}(1/N^2)$. 

\subsection{Bubble resummation}

The knowledge of the critical exponent 
$\omega_-$ at $\mathcal{O}(1/N)$
is enough to obtain the explicit form of 
$\beta_y$ in Eq.~\eqref{eq:beta1}
at the same order in $1/N$.
This is not the case for $\beta_{\lambda}$ in Eq.~\eqref{eq:beta2},
as the information contained in
$\omega_+$ can only constrain
a combinations of $F_1$, $F_3$ and $F_4$, see Eq.~\eqref{eq:omplus}.
In order to obtain $\beta_{\lambda}$ at the order $1/N$,
we have thus to rely on explicit bubble resummation.

The $\beta$-function for $\lambda$ is obtained
by acting with derivatives on the 1PI vertex
counterterm, $Z_\lambda$, and on the scalar
self-energy counterterm, $Z_S$.
The bare coupling, $\lambda_0$, and the renormalized
coupling, $\lambda$, are related via
\begin{equation}\label{eq:g2r}
\lambda_0 = Z_{\lambda} Z_S^{-2} \lambda,
\end{equation}
and the $\beta$-function is
\begin{equation}\label{eq:beta}
\beta_\lambda = 
\lambda \left( \lambda \frac{\partial}{\partial \lambda}
+ K \frac{ \partial}{\partial K}\right) \, \text{ln}\, \left( Z_\lambda Z_S^{-2}\right)_{1/\epsilon}.
\end{equation}
The self-energy renormalisation constant, $Z_S$, has been computed in Ref.~\cite{Alanne:2018ene}
up to $\mathcal{O}(1/N)$, and reads:
\begin{equation}\label{eq:zetaS}
 Z_S = 1 - \frac{K}{\epsilon} - \frac{1}{ \epsilon N} 
 \int_0^K \left( \xi_0(t) - \xi_0(0) + \xi(t,1) K \right) \mathrm{d}t,
\end{equation}
where 
\begin{equation}
\xi_0(t)=-\frac{(1-t)\Gamma(4-t)}{\Gamma\left(2-\frac{t}{2}\right)
    \Gamma\left(3-\frac{t}{2}\right)\pi t}\sin\left(\frac{\pi t}{2}\right),
\end{equation}
and
\begin{equation}
 \xi(t,1) = \frac{1}{1-t} \xi_0(t).
\end{equation}
The coupling-constant renormalisation constant, $Z_\lambda$, is given by
\begin{equation}\label{eq:Zlambda}
Z_\lambda = 1 - \text{div}\{ Z_\lambda \Lambda_0(\lambda_0, K_0, p^2, \epsilon)\},
\end{equation}
where $K_0$ is the rescaled Yukawa coupling,
and $\Lambda_0$ contains 
the 1PI contributions to the four-point function. 

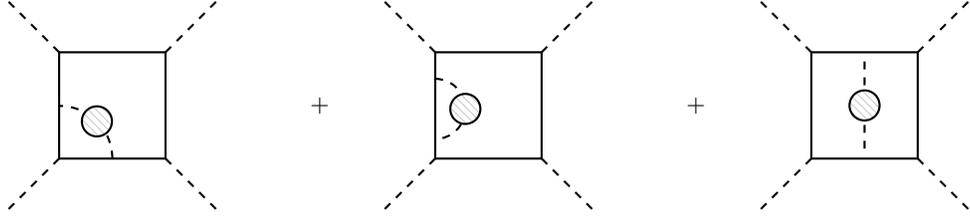
\begin{figure*}[t]
    \centering
  \begin{tikzpicture}[node distance=1cm]
    \coordinate (v1);
    \coordinate[above left = of v1] (v1a);
    \coordinate[below left = of v1] (v1b);
    \coordinate[above right = of v1] (v1c);
    \coordinate[below right = of v1] (v1d);
    \coordinate[above left = of v1a] (va);
    \coordinate[below left = of v1b] (vb);
    \coordinate[above right = of v1c] (vc);
    \coordinate[below right = of v1d] (vd);
    \draw[rscalar] (v1a) -- (va);
    \draw[rscalar] (v1b) -- (vb);
    \draw[rscalar] (v1c) -- (vc);
    \draw[rscalar] (v1d) -- (vd);
    \draw[mfermion] (v1b) -- (v1d)
	node[pos=0.5] (v1d2) {};
    \draw[mfermion] (v1d)-- (v1c) -- (v1a) -- (v1b);
    \draw[rscalar] (v1d2) arc(0:90:0.7)
	node[pos=0.5,solid,whiteblob2,minimum size=0.4 cm] {}
	node[pos=0.5,draw,solid,blob,minimum size=0.4 cm] {};
    \coordinate[right = of v1] (v01);
    \coordinate[right = of v01] (v012);
    \coordinate[right = of v012,label= left:$+$] (vA);
    \coordinate[right = of vA] (v0A);
    \coordinate[right = of v0A] (v2);
    \coordinate[above left = of v2] (v2a);
    \coordinate[below left = of v2] (v2b);
    \coordinate[above right = of v2] (v2c);
    \coordinate[below right = of v2] (v2d);
    \coordinate[above left = of v2a] (va2);
    \coordinate[below left = of v2b] (vb2);
    \coordinate[above right = of v2c] (vc2);
    \coordinate[below right = of v2d] (vd2);
    \draw[rscalar] (v2a) -- (va2);
    \draw[rscalar] (v2b) -- (vb2);
    \draw[rscalar] (v2c) -- (vc2);
    \draw[rscalar] (v2d) -- (vd2);
    \draw[mfermion] (v2a) -- (v2b)
	node[pos=0.25] (v2a2) {};
    \draw[mfermion] (v2b) -- (v2d)-- (v2c) -- (v2a);
    \draw[rscalar] (v2a2) arc(90:-90:0.4)
	node[pos=0.5,solid,whiteblob2,minimum size=0.4 cm] {}
	node[pos=0.5,draw,solid,blob,minimum size=0.4 cm] {};
    \coordinate[right = of v2] (v02);
    \coordinate[right = of v02] (v022);
    \coordinate[right = of v022,label= left:$+$] (vB);
    \coordinate[right = of vB] (v0B);
    \coordinate[right = of v0B] (v3);
    \coordinate[above left = of v3] (v3a);
    \coordinate[below left = of v3] (v3b);
    \coordinate[above right = of v3] (v3c);
    \coordinate[below right = of v3] (v3d);
    \coordinate[above left = of v3a] (va3);
    \coordinate[below left = of v3b] (vb3);
    \coordinate[above right = of v3c] (vc3);
    \coordinate[below right = of v3d] (vd3);
    \draw[rscalar] (v3a) -- (va3);
    \draw[rscalar] (v3b) -- (vb3);
    \draw[rscalar] (v3c) -- (vc3);
    \draw[rscalar] (v3d) -- (vd3);
    \draw[mfermion] (v3a) -- (v3b);
    \draw[mfermion] (v3b) -- (v3d)
	node[pos=0.5] (v3d2) {};
    \draw[mfermion] (v3d) -- (v3c);
    \draw[mfermion] (v3c) -- (v3a)
	node[pos=0.5] (v3a2) {};
    \draw[rscalar] (v3a2) -- (v3d2)
	node[pos=0.5,solid,whiteblob2,minimum size=0.4 cm] {}
	node[pos=0.5,draw,solid,blob,minimum size=0.4 cm] {};
  \end{tikzpicture}
    \caption{Different types of box diagrams for $F_2(y N)$. The gray blob represents a chain of fermion bubbles.}
  \label{fig:box}
\end{figure*}

At the order $\mathcal{O}(1/N)$, we have
\begin{equation}\label{eq:Lambda0}
\begin{split}
\Lambda_0 = & \lambda_0 \sum_{n=0}^\infty K_0^n \Lambda^{(n+1)}_\lambda(p^2,\epsilon) + 
\frac{1}{\lambda_0 N} K_0^2 \Lambda_K^{(1)}(p^2,\epsilon) \\
&+\frac{1}{\lambda_0 N^2} K_0^3 \sum_{n=0}^\infty K_0^n \Lambda_K^{(n+2)}(p^2,\epsilon)\\
&+ \frac{1}{N} K_0^2 \sum_{n=0}^\infty K_0^n L^{(n+2)}(p^2,\epsilon) \\
&+ \frac{1}{\lambda_0 N^2} K_0^4 \sum_{n=0}^\infty K_0^n {\Lambda_K^\prime}^{(n+3)}(p^2,\epsilon).
\end{split}
\end{equation}
The first term corresponds to a basic candy diagram where the Yukawa couplings only enter
through the chain of fermion bubbles. The second term is the basic one-loop box diagram.
The third term is a box diagram with an additional internal scalar propagator.
The fourth term is a candy with two different vertices, namely one $\lambda$ and one
effective quartic made of a fermion loop. The last term is a three-loop candy diagram
with two fermion loops as effective quartics. The different topologies are shown in 
Figs~\ref{fig:cc}--\ref{fig:candies}. 

The $p^2$ in the arguments refers generically to the IR regulator. 
We use two IR regulation strategies depending on the subclass of diagrams: For fermion-box type diagrams, we 
use a convenient choice of non-zero external momenta, and for the scalar-candy-type diagrams we give a non-zero regulating mass for the 
propagating scalars. The sum of the contributions in 
each of these subclasses is IR finite, justifying the different regularisations.

After trading the bare couplings with the renormalized couplings,
\begin{equation}
\lambda_0 = Z_\lambda Z_S^{-2} \lambda, \quad K_0 = Z_S^{-1}(Z_V Z_F^{-1})^2 K,
\end{equation}
where $Z_{V,F}=1+\mathcal{O}(1/N)$ are the renormalisation constants for the 1PI Yukawa vertex and fermion self-energy, resp., and keeping only term that contribute up to $\mathcal{O}(1/N)$\footnote{We assume here $\lambda \sim 1/N$.}, 
we find for~$Z_\lambda$:
\begin{align}\label{eq:zetalambda}
Z_\lambda = 1 &- \text{div}\bigg\{\lambda Z_S^{-2}\sum_{n=0}^\infty (Z_S^{-1} K)^n \Lambda_\lambda^{(n+1)} \nonumber\\ &
+ \frac{1}{N} K^2 Z_S^{-2} 
\sum_{n=0}^\infty (Z_S^{-1}K)^n \left(L^{(n+2)}-2 D \Lambda_\lambda^{(n+1)}\right)\nonumber\\
&+\frac{1}{\lambda N} K^2 (Z_V Z_F^{-1})^4 \Lambda_K^{(1)}\nonumber\\
&+\frac{1}{\lambda N^2} K^3 Z_S^{-1} \sum_{n=0}^\infty (Z_S^{-1}K)^n \Lambda_K^{(n+2)}\\ 
&+ \frac{1}{\lambda N^2} K^4 Z_S^{-2} \sum_{n=0}^\infty (Z_S^{-1}K)^n 
\left( D^2{\Lambda_\lambda}^{(n+1)}\right.\nonumber\\
&\left.\qquad\qquad- D {L}^{(n+2)}+ {\Lambda_K^\prime}^{(n+3)}\right)
\bigg\}\nonumber,
\end{align}
where we have iterated Eq.~\eqref{eq:Zlambda} to include all 
the contributions up to $\mathcal{O}(1/N)$, and we
have defined $D$ as
\begin{equation}
D = \text{div}\{ \Lambda_K^{(1)}\}.
\end{equation}
Notice that, despite the explicit $1/N^2$ dependence,
the $1/\lambda N^2$ contributions are actually $\mathcal{O}(1/N)$ when 
interpreted in terms of the rescaled quartic, $\lambda N$.

After taking derivatives according to Eq.~\eqref{eq:beta}, 
the first line in Eq.~\eqref{eq:zetalambda} will give the $\lambda^2$-contribution 
in Eq.~\eqref{eq:beta2}, namely the function $F_3$,
and the second line will contribute to $F_4$.
The last four lines behave like $1/\lambda$ and 
correspond to the pure Yukawa contribution, $F_2$.

\subsection{$F_3$, $F_4$ and cross-check}
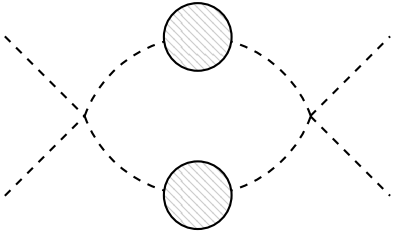
\begin{figure}[t]
\begin{tikzpicture}[node distance=1.5cm]
    \coordinate (v1);
    \coordinate[above left = of v1] (va);
    \coordinate[below left = of v1] (vb);
    \coordinate[right = of v1] (v2);
    \coordinate[right = of v2] (v3);
    \coordinate[above right = of v3] (vc);
    \coordinate[below right = of v3] (vd);
    \draw[rscalar] (va) -- (v1);
    \draw[rscalar] (vb) -- (v1);
    \draw[rscalar] (v1) arc(160:20:1.6) 
	node[pos=0.5,solid,whiteblob2,minimum size=0.9 cm] {}
	node[pos=0.5,draw,solid,blob,minimum size=0.9 cm] {};
    \draw[rscalar] (v1) arc(-160:-20:1.6)
	node[pos=0.5,solid,whiteblob2,minimum size=0.9 cm] {}
	node[pos=0.5,draw,solid,blob,minimum size=0.9 cm] {};
    \draw[rscalar] (v3) -- (vc);
    \draw[rscalar] (v3) --  (vd); 
  \end{tikzpicture}
    \caption{Diagram for $F_3(y N)$. The gray blob represents a bubble chain. }
    \label{fig:cc} 
\end{figure}

We compute here $F_3$ and $F_4$ by explicit resummation
and cross-check our result with Eq.~\eqref{eq:omplus}.
Diagrammatically, $F_3$ corresponds to 
Fig.~\ref{fig:cc}, where the internal scalar lines
are dressed with fermion bubbles.
To obtain its expression, we refer to the first 
line
of Eq.~\eqref{eq:zetalambda} and compute:
\begin{align}
 T_3 \equiv - & \text{div}\left\{ 
\lambda Z_S^{-2}\sum_{n=0}^\infty (Z_S^{-1} K)^n \Lambda_\lambda^{(n+1)}\right\} \\
 = - &\text{div} \left\{ \lambda \sum_{n=0}^\infty K^n 
 \sum_{i=0}^n \left(\begin{array}{c} n+1 \\ i \end{array} \right)\frac{(-1)^i}{\epsilon^i} \Lambda_\lambda^{(n-i+1)} \right\}.\nonumber
\end{align}
The function $\Lambda_\lambda^{(m)}$ is found to be
\begin{equation}
\Lambda_\lambda^{(m)} = \frac{1}{\epsilon^m} l(\epsilon,m)
= \frac{1}{\epsilon^m} \sum_{j=0}^\infty (m \epsilon)^j l_j(\epsilon).
\end{equation}
After resummation, and keeping only the $1/\epsilon$ pole of $T_3$, 
we find
\begin{equation}
T_3 = - \frac{1}{\epsilon} \lambda \, l_0(K)\ 
+\dots\, ,
\end{equation}
where
\begin{equation}
l_0(t) = \frac{ 9\cdot 2^{4-t} \Gamma \left( \frac{3}{2} 
- \frac{t}{2} \right) \text{sin} \left( \frac{ \pi t}{2}\right)}
{\pi^{3/2} t \Gamma \left( 2 - \frac{t}{2}\right) }.
\end{equation}
By adopting the same convention of
Eq.~\eqref{eq:beta2}, the function $F_3$ is  
\begin{equation}\label{eq:F3}
 F_3(t) = l_0(2 t) +2 t l_0^\prime(2 t) - 36,
\end{equation}
where $-36$ removes the one-loop contribution.

To compute $F_4$, we start from the second 
line
of Eq.~\eqref{eq:zetalambda}, which diagrammatically corresponds 
to Fig.~\ref{fig:L}:
\begin{figure}[t]
  \centering
\begin{minipage}[c]{0.48\textwidth}
\centering
\begin{tikzpicture}[node distance=1cm]
    \coordinate[label = center:{\Large\ \ $N,T$}] (v1);
    \coordinate[right = of v1] (v2);
    \coordinate[right = of v2] (v3);
    \coordinate[right = of v3] (v4);
    \coordinate[above right = of v4] (vc);
    \coordinate[below right = of v4] (vd);
    \draw[mfermion] (v2) arc(0:45:0.8) coordinate (v1c) arc(45:135:0.8) coordinate (v1a)
	arc(135:225:0.8) coordinate (v1b) arc(225:315:0.8) coordinate (v1d) arc(315:360:0.8);
    \coordinate[above left = of v1a] (va);
    \coordinate[below left = of v1b] (vb);
    \draw[rscalar] (va) -- (v1a);
    \draw[rscalar] (vb) -- (v1b);
    \draw[rscalar] (v1c) arc(120:30:1.6) 
	node[pos=0.5,solid,whiteblob2,minimum size=0.9 cm] {}
	node[pos=0.5,draw,solid,blob,minimum size=0.9 cm] {};
    \draw[rscalar] (v1d) arc(-120:-30:1.6)
	node[pos=0.5,solid,whiteblob2,minimum size=0.9 cm] {}
	node[pos=0.5,draw,solid,blob,minimum size=0.9 cm] {};
    \draw[rscalar] (v4) -- (vc);
    \draw[rscalar] (v4) --  (vd); 
  \end{tikzpicture}
  \end{minipage}\\
  \vspace{0.5cm} 
  \begin{minipage}[c]{0.22\textwidth}
	\begin{tikzpicture}[node distance=0.6cm]
	    \coordinate[label = center: {\Large $N$}] (v1);
	    \draw (v1) circle [radius=0.5cm];
	    \coordinate[right = of v1,label = right:{\Large $=$}] (v2);
	    \coordinate[right = of v2] (v3);
	    \coordinate[right = of v3] (v4);
	    \coordinate[right = of v4] (v5);
	    \draw[mfermion] (v5) arc(0:45:0.5) coordinate (v1c)  arc(45:135:0.5) coordinate (v1a)
		arc(135:225:0.5) coordinate (v1b) arc(225:315:0.5) coordinate (v1d) arc(315:360:0.5);
	    \draw[fill=black] (v1a) circle [radius=2pt];
	    \draw[fill=black] (v1b) circle [radius=2pt];
	    \draw[fill=black] (v1c) circle [radius=2pt];
	    \draw[fill=black] (v1d) circle [radius=2pt];
	\end{tikzpicture}
  \end{minipage}\quad
  \begin{minipage}[c]{0.22\textwidth}
	\centering
	\begin{tikzpicture}[node distance=0.6cm]
	    \coordinate[label = center: {\Large $T$}] (v1);
	    \draw (v1) circle [radius=0.5cm];
	    \coordinate[right = of v1,label = right:{\Large $=$}] (v2);
	    \coordinate[right = of v2] (v3);
	    \coordinate[right = of v3] (v5);
	    \coordinate[above left = of v5] (v5a);
	    \coordinate[below left = of v5] (v5b);
	    \coordinate[above right = of v5] (v5c);
	    \coordinate[below right = of v5] (v5d);
	    \draw[color=white] (v5a) -- (v5d)
		node[pos=0.345] (v6a) {}
		node[pos=0.48] (v6) {}
		node[pos=0.54] (v7) {};
	    \draw[mfermion] (v5a) -- (v5c) -- (v5b) -- (v5d);
	    \draw[mfermion] (v5a) -- (v6); 
	    \draw[mfermion] (v7) -- (v5d);
	    \draw[mfermion] (v6a) arc(137:-45:0.2);
	    \draw[fill=black] (v5a) circle [radius=2pt];
	    \draw[fill=black] (v5b) circle [radius=2pt];
	    \draw[fill=black] (v5c) circle [radius=2pt];
	    \draw[fill=black] (v5d) circle [radius=2pt];
	\end{tikzpicture}
  \end{minipage}
    \caption{Diagram for $F_4(y N)$. The gray blob represents a chain of fermion bubbles. The labels $N$ and $T$ correspond to 
    non-twisted and twisted fermion bubbles, respectively, and are pictorially represented below the $F_4$ diagram. 
    }
    \label{fig:L}
\end{figure}
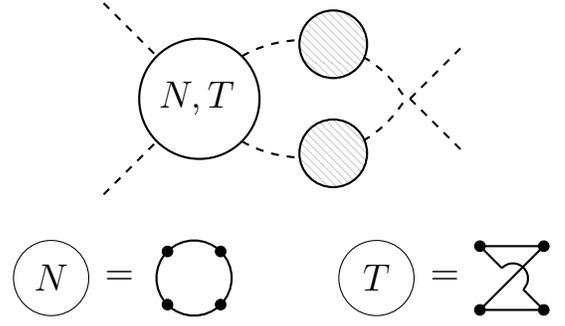
\begin{align}
&T_4 \equiv - \frac{K^2}{N} \text{div}\left\{
Z_S^{-2} \sum_{n=0}^\infty (Z_S^{-1}K)^n \left(L^{(n+2)}-2 D \Lambda_\lambda^{(n+1)}\right) \right\} \nonumber\\
&= - \frac{K^2}{N} \text{div}\left\{
\sum_{n=0}^\infty  K^n \sum_{i=0}^{n} 
\left( \begin{array}{c} n+1 \\ i \end{array} \right)
\frac{(-1)^i}{\epsilon^i} \Gamma(n+2-i,\epsilon) \right\}
\end{align}
where we have introduced
\begin{equation}
\Gamma(m,\epsilon) \equiv L^{(m)}-2 D \Lambda_\lambda^{(m-1)} = \frac{1}{m \epsilon^m} (-1)^m \gamma(\epsilon,m),
\end{equation}
with
\begin{equation}
\gamma(\epsilon,m) = \sum_{j=0}^\infty (m \epsilon)^j \gamma_j(\epsilon).
\end{equation}
After resummation, and keeping only the $1/\epsilon$ pole of $T_4$, 
we find
\begin{equation}
T_4 = \frac{1}{\epsilon N} \left(K \gamma(K,1)-
\int_0^K \gamma_0(t) \text{d}t \right) 
+\dots\, ,
\end{equation}
where
\begin{equation}
\gamma_0(t) = \frac{3\cdot2^{3-t}(t-3)\Gamma\left( \frac{3}{2}-\frac{t}{2}\right)}{
\pi^{3/2} t \Gamma(2 - t/2)} \text{sin} \left( \frac{\pi t}{2}\right),
\end{equation}
and
\begin{equation}
\gamma(t,1) = \frac{1}{1-t/3} \gamma_0(t).
\end{equation}
%
Besides $T_4$, the function $F_4$ gets contribution from 
$Z_S$ in Eq.~\eqref{eq:zetaS}. Altogether,
in the convention of Eq.~\eqref{eq:beta2}, it is given by
\begin{align}\label{eq:F4}
 F_4(t) = & \,
 2 \gamma(2 t,1) - 2 \gamma_0(2 t) + 
 4 t \gamma'(2t,1) \\
 &
 + 4 \xi_0(2t) + 8 t \xi(2t,1) +6 
 + 4 \int_0^{2 t} \xi(x,1) \text{d}x.\nonumber
\end{align}

With the results of Eqs~\eqref{eq:F3} and~\eqref{eq:F4} together with Eq.~\eqref{eq:F1GNY},
one can check that Eq.~\eqref{eq:omplus} is fulfilled. This provides a powerful 
cross-check for our computation.

\subsection{The function $F_2$}
The function $F_2$ can be computed evaluating the $1/\lambda$ terms in the last four 
lines of Eq.~\eqref{eq:zetalambda}, 
corresponding to the one-loop box diagram, the box diagrams with additional internal scalar propagator
in Fig.~\ref{fig:box} and to the candy diagrams in
Fig.~\ref{fig:candies}.

Let us start with the box contribution.
The counterterms $Z_V$ and $Z_F$ have been computed 
up to $\mathcal{O}(1/N)$ in Ref.~\cite{Alanne:2018ene}
and read:
\begin{equation}\label{eq:ZVF}
\begin{split}
& Z_V = 1 + \frac{1}{2 N} 
\sum_{n=1}^\infty \frac{K^n}{\epsilon^n} \frac{v_0(\epsilon)}{n} \\
& Z_F = 1 - \frac{1}{4 N} 
\sum_{n=1}^\infty \frac{K^n}{\epsilon^n} \frac{\sigma_0(\epsilon)}{n}.
\end{split}
\end{equation}
The third 
line in Eq.~\eqref{eq:zetalambda} 
gives a divergent part
\begin{equation}\label{eq:T1}
T_{b_1} \equiv
\frac{4K^2}{\lambda N} \left[ \frac{D}{4}
+ \text{div} \left\{(\tilde{Z}_F - \tilde{Z}_V) \Lambda_F^{(1)}(p^2,\epsilon)
\right\}\right],
\end{equation}
where $\tilde{Z}_{F,V} \equiv Z_{F,V} - 1$, and we have defined the finite part of the one-loop box diagram as
\begin{equation}
 \Lambda_F^{(1)} = \Lambda^{(1)}_K -  D.
\end{equation}
The first term in Eq.~\eqref{eq:T1} gives the basic one-loop
contribution of the box diagram and will be omitted in the following.
Using Eq.~\eqref{eq:ZVF} and keeping only the $1/\epsilon$ pole of $T_{b_1}$,
we have:
\begin{equation}
T_{b_1} = 
\frac{K^2}{\epsilon \lambda N^2} \int_0^K \Lambda_F^{(1)}(p^2,t)(\sigma_0(t) + 2 v_0(t)) \dd t\ 
+\dots
\end{equation}
As for the fourth line in Eq.~\eqref{eq:zetalambda}, 
we have 
\begin{equation}\label{eq:T2}
T_{b_2}\equiv\frac{K^3}{\lambda N^2} \sum_{n=0}^\infty K^n \text{div} \sum_{i=0}^n 
\left\{ \left( \begin{array}{c} n \\ i \end{array} \right) 
\frac{1}{\epsilon^i} \Lambda^{(n+2-i)}_K (p^2,\epsilon) \right\}.
\end{equation}
The quantity $\Lambda^{(m)}_K$ allows for the following expansion:
\begin{equation}\label{eq:Lambdam}
\Lambda^{(m)}_K(p^2,\epsilon) = \frac{1}{\epsilon^m m(m-1)}\lambda(p^2,\epsilon,m),
\end{equation}
where $\lambda(p^2,\epsilon,m)$ is regular for $\epsilon \rightarrow 0$ and 
can be written as
\begin{equation}\label{eq:lambdam}
 \lambda(m,\epsilon) = \sum_{j=0}^\infty (m \epsilon)^j \lambda_j(p^2,\epsilon).
\end{equation}
Plugging Eqs~\eqref{eq:Lambdam} and \eqref{eq:lambdam} in Eq.~\eqref{eq:T2}
and using the usual summation formulas, we find
for the $1/\epsilon$ pole:
\begin{equation}
\begin{split}
T_{b_2} = \frac{K}{\epsilon \lambda N^2}
 &\int_0^K \left(\vphantom{\frac12} \lambda_0(t) -\lambda_0(0) \right.\\
 &\left.\qquad\ +K \frac{\lambda(p^2,t,1) - \lambda_0(t)}{t} \right) \text{d}t
\ +\dots
\end{split}
\end{equation}

When the $1/\epsilon$ poles of $T_{b_1}$ and $T_{b_2}$ are put together,
the $p^2$ dependence of $\lambda(p^2,t,1)$ 
cancels.
We find the $1/\epsilon$ pole of $Z_\lambda^\text{box}$ to be
\begin{equation}\begin{split}
Z_\lambda^\text{box} & = -( T_{b_1}+T_{b_2})\\
& = -\frac{K}{\epsilon \lambda N^2} \int_0^K \left(\vphantom{\frac12} \lambda_0(t) -\lambda_0(0)\right.\\
&\left.\qquad\qquad\qquad\quad+ K \frac{5-t}{4-5 t+t^2} \lambda_0(t) \right) \text{d}t,
\end{split}
\end{equation}
where the function $\lambda_0(t)$ is given by
\begin{equation}
\lambda_0(t) = (t-1) \frac{ \Gamma(4-t)\, \text{sin}\left( \frac{ \pi t}{2} \right)}
{\pi t \Gamma\left(2- \frac{t}{2}\right)^2}.
\end{equation}
In the convention of Eq.~\eqref{eq:beta2}, 
\begin{equation}\begin{split}\label{eq:F2box}
 F_2^\text{box}(t) = 
 &
\int_0^{2t} \frac{x-5}{x^2-5x+4} \lambda_0(x) \text{d}x \\
&+\lambda_0(0)- \frac{4}{4-10t+4t^2}\lambda_0(2 t). 
\end{split}
\end{equation}

Let us now compute the contribution of the candy diagrams 
in Fig.~\ref{fig:candies}, $Z_\lambda^\text{candy}$,
arising from the last two lines of Eq.~\eqref{eq:zetalambda}.
It can be rewritten as
\begin{equation}\label{eq:Tc}
~\hspace{-0.1cm}T_c \equiv \frac{K^4}{\lambda N^2}\mathrm{div}\left\{ \sum_{n=0}^\infty K^n
\sum_{i=0}^n
\left(\hspace{-0.1cm} \begin{array}{c} n + 1 \\ i \end{array}\hspace{-0.1cm} \right) 
\frac{1}{\epsilon^i} C^{(n+1-i)}\right\},
\end{equation}
where 
\begin{equation}
C^{(n+1-i)}\equiv D^2{\Lambda_\lambda}^{(n+1-i)}
- D {L}^{(n+2-i)}
+ {\Lambda_K^{\prime\,(n+3-i)}}.
\end{equation}
\begin{figure}[t]
    \centering
\begin{minipage}[c]{0.48\textwidth}
\begin{tikzpicture}[node distance=1cm]
    \coordinate[label = center:{\Large\ \ $A$}] (v1);
    \coordinate[left = 1.6 cm of v1, label = left:{\Large $\Lambda_K^{\prime}$:}] (v0);
    \coordinate[right = of v1] (v2);
    \coordinate[right = of v2] (v3);
    \coordinate[right = of v3,label = center:{\Large\ \ $B$}] (v4);
    \coordinate[right = of v4] (v6);
    \draw[mfermion] (v2) arc(0:45:0.8) coordinate (v1c) arc(45:135:0.8) coordinate (v1a)
	arc(135:225:0.8) coordinate (v1b) arc(225:315:0.8) coordinate (v1d) arc(315:360:0.8);
    \coordinate[above left = of v1a] (va);
    \coordinate[below left = of v1b] (vb);
    \draw[rscalar] (va) -- (v1a);
    \draw[rscalar] (vb) -- (v1b);
    \draw[rscalar] (v1c) arc(120:60:2) 
	node[pos=0.5,solid,whiteblob2,minimum size=0.9 cm] {}
	node[pos=0.5,draw,solid,blob,minimum size=0.9 cm] {};
    \draw[rscalar] (v1d) arc(-120:-60:2)
	node[pos=0.5,solid,whiteblob2,minimum size=0.9 cm] {}
	node[pos=0.5,draw,solid,blob,minimum size=0.9 cm] {};
    \draw[mfermion] (v6) arc(0:45:0.8) coordinate (v6c) arc(45:135:0.8) coordinate (v6a)
	arc(135:225:0.8) coordinate (v6b) arc(225:315:0.8) coordinate (v6d) arc(315:360:0.8);
    \coordinate[above right = of v6c] (vc);
    \coordinate[below right = of v6d] (vd);
    \draw[rscalar] (vd) -- (v6d);
    \draw[rscalar] (vc) -- (v6c);
  \end{tikzpicture}
  \end{minipage}\\
  \vspace{0.5cm}
\begin{minipage}[c]{0.48\textwidth}
\begin{tikzpicture}[node distance=1cm]
    \coordinate[label = center:{\Large\ \ $A$}] (v1);
    \coordinate[left = 1.4cm of v1, label = left:{\Large$D L:$}] (v0);
    \coordinate[right = of v1] (v2);
    \coordinate[right = of v2] (v3);
    \coordinate[right = 1.4cm of v3] (v4);
    \coordinate[above right = 1.5cm of v4] (vc);
    \coordinate[below right = 1.5cm of v4] (vd);
    \draw[mfermion] (v2) arc(0:45:0.8) coordinate (v1c) arc(45:135:0.8) coordinate (v1a)
	arc(135:225:0.8) coordinate (v1b) arc(225:315:0.8) coordinate (v1d) arc(315:360:0.8);
    \coordinate[above left = of v1a] (va);
    \coordinate[below left = of v1b] (vb);
    \draw[rscalar] (va) -- (v1a);
    \draw[rscalar] (vb) -- (v1b);
    \draw[rscalar] (v1c) arc(120:30:2) 
	node[pos=0.35,solid,whiteblob2,minimum size=0.9 cm] {}
	node[pos=0.35,draw,solid,blob,minimum size=0.9 cm] {};
    \draw[rscalar] (v1d) arc(-120:-30:2)
	node[pos=0.35,solid,whiteblob2,minimum size=0.9 cm] {}
	node[pos=0.35,draw,solid,blob,minimum size=0.9 cm] {};
    \draw[rscalar] (v4) -- (vc);
    \draw[rscalar] (v4) --  (vd); 
    \draw[fill=white,thick] (v4) circle [radius=5pt]
	node {{\Large$\times$}};
  \end{tikzpicture}
  \end{minipage}\\
  \vspace{0.5cm}
\begin{minipage}[c]{0.48\textwidth}
\begin{tikzpicture}[node distance=1.5cm]
    \coordinate (v1);
    \coordinate[left =1.3cm of v1, label = left:{\Large$D^2\Lambda_{\lambda}:$}] (v0);
    \coordinate[above left = 1.8cm of v1] (va);
    \coordinate[below left = 1.8cm of v1] (vb);
    \coordinate[right = of v1] (v2);
    \coordinate[right = of v2] (v3);
    \coordinate[above right = 1.7cm of v3] (vc);
    \coordinate[below right = 1.7cm of v3] (vd);
    \draw[rscalar] (va) -- (v1);
    \draw[rscalar] (vb) -- (v1);
    \draw[rscalar] (v1) arc(160:20:1.6) 
	node[pos=0.5,solid,whiteblob2,minimum size=0.9 cm] {}
	node[pos=0.5,draw,solid,blob,minimum size=0.9 cm] {};
    \draw[rscalar] (v1) arc(-160:-20:1.6)
	node[pos=0.5,solid,whiteblob2,minimum size=0.9 cm] {}
	node[pos=0.5,draw,solid,blob,minimum size=0.9 cm] {};
    \draw[rscalar] (v3) -- (vc);
    \draw[rscalar] (v3) --  (vd); 
    \draw[fill=white,thick] (v1) circle [radius=5pt]
	node {{\Large$\times$}};
    \draw[fill=white,thick] (v3) circle [radius=5pt]
	node {{\Large$\times$}};
  \end{tikzpicture}
  \end{minipage}
    \caption{Candy diagrams for $F_2(y N)$.
    The basic diagram is the three-loop diagram $\Lambda^\prime_K$. The fermion loops $A,B$ 
are either of the form $N$ (non-twisted) or $T$ (twisted), cf. Fig.~\ref{fig:L}.
The diagrams $D L$ and $D^2 \Lambda_\lambda$ come from $\Lambda^\prime_K$ when $A$ or $B$ are
shrunk to a counterterm
vertex $D$, which is the same for $N$ and $T$.
}
    \label{fig:candies}
\end{figure}
The structure of $C^{(m)}$ is:
\begin{equation}\label{eq:Cm}
C^{(m)} = \frac{1}{\epsilon^{m+2}} \frac{1}{(m+1)(m+2)} c(p^2,\epsilon,m),
\end{equation}
where the function $c(p^2,\epsilon,m)$ is regular for $\epsilon \rightarrow 0$
and can be expanded as
\begin{equation}\label{eq:cm}
c(p^2,\epsilon,m) =
\sum_{j=0}^\infty (m \epsilon)^j c_j(p^2,\epsilon).
\end{equation}
The IR regulator, $p^2$, stands here for a soft mass
for the scalar field.
Plugging Eqs~\eqref{eq:Cm} and \eqref{eq:cm} in Eq.~\eqref{eq:Tc} yields
\begin{equation}
 T_c = \frac{K^4}{\lambda N^2}\mathrm{div}\left\{ \sum_{n=0}^\infty 
 \frac{K^n}{\epsilon^{n+3}} \sum_{j=0}^{n+2} \epsilon^j c_j(p^2,\epsilon) 
 S(n,j)\right\},
\end{equation}
where
\begin{equation}
 S(n,j) = \sum_{i=0}^n
\left( \begin{array}{c} n + 1 \\ i \end{array} \right) (-1)^i
\frac{(n+2-i)^{j-1}}{n+3-i}.
\end{equation}
We find
\begin{equation}\label{eq:S}
S(n,j)
= \begin{cases}
(-1)^n \frac{n+1}{2 (n+3)} \quad & j=0, \\
(-1)^n \frac{(n+1)(n+4)}{2(n+2)(n+3)} & j \, \text{odd}, \\
(-1)^n \frac{ 8 + n (n+5)}{2 (n+2)(n+3)} & j \, \text{even}.
\end{cases}
\end{equation}
Eq.\,\eqref{eq:S} tells that three functions are relevant: $c_0(\epsilon)$,
$c_e(\epsilon)$, and $c_o(\epsilon)$,
\begin{align}
 c_e(\epsilon) &\equiv \sum_{j \, \text{even}}^\infty \epsilon^j c_j(\epsilon)
 = \frac{c(p^2,\epsilon,1)+c(p^2,\epsilon,-1)}{2} -c_0(\epsilon),\nonumber\\
 &~\\
c_o(\epsilon) &\equiv 
\sum_{j \, \text{odd}}^\infty \epsilon^j c_j(\epsilon)
=  \frac{c(p^2,\epsilon,1)-c(p^2,\epsilon,-1)}{2},\nonumber
\end{align}
where the summation over $j$ has been extended to $\infty$
without affecting the result (namely, finite terms in the 
$\epsilon \rightarrow 0$ limit),  and
all the resulting functions are found to be independent of the IR regulator. 
They read
\begin{align}
c_0(t) =& -3\frac{ 2^{4-t} \,
\Gamma\left( \frac{5}{2} - \frac{t}{2}\right)
\text{sin}\left( \frac{\pi t}{2}\right)}
{ \pi^{3/2} t\Gamma\left(2-\frac{t}{2}\right)},\\
c_e(t) =& \frac{t^2}{6(3 - t)} c_0(t),\\
c_o(t) =& \frac{ t(6-t)}{6(3-t)} c_0(t). \label{eq:co}
\end{align}
After performing the sum over $n$ using Eqs~\eqref{eq:S}--\eqref{eq:co},
and retaining only the $1/\epsilon$ pole, we arrive at
\begin{align}
T_c = \frac{1}{\epsilon \lambda N^2}&
 \bigg( \frac{1}{2} K^2 \left(c_0(K) + c_0(0)\right)
 + \frac{K^3}{6-2K}c_0(K)\nonumber \\ 
 & +\frac{1}{3} K \int_0^K (t-3-K) c_0(t) \text{d}t\bigg)\ +\dots
\end{align}
In the convention of Eq.~\eqref{eq:beta2}, 
\begin{align}\label{eq:F2cand}
 F_2^\text{candy}(t) = 
 &\,\frac{9-24 t + 8 t^2 }{2(2t-3)^2}c_0(2 t) +\frac{3 t}{2t-3}c_0^\prime(2t) \nonumber\\ &
 -\frac{1}{6} \left( 3 c_0(0) - 2 \int_0^{2t} c_0(x)\text{d}x \right). 
\end{align}
Combining Eqs~\eqref{eq:F2box} and~\eqref{eq:F2cand}, the function $F_2(t)$ is then
\begin{equation}
\label{eq:F2}
 F_2(t) = F_2^\text{box}(t) + F_2^\text{candy}(t).
\end{equation}

\begin{figure}[t]
    \begin{center}
	\includegraphics[width=0.48\textwidth]{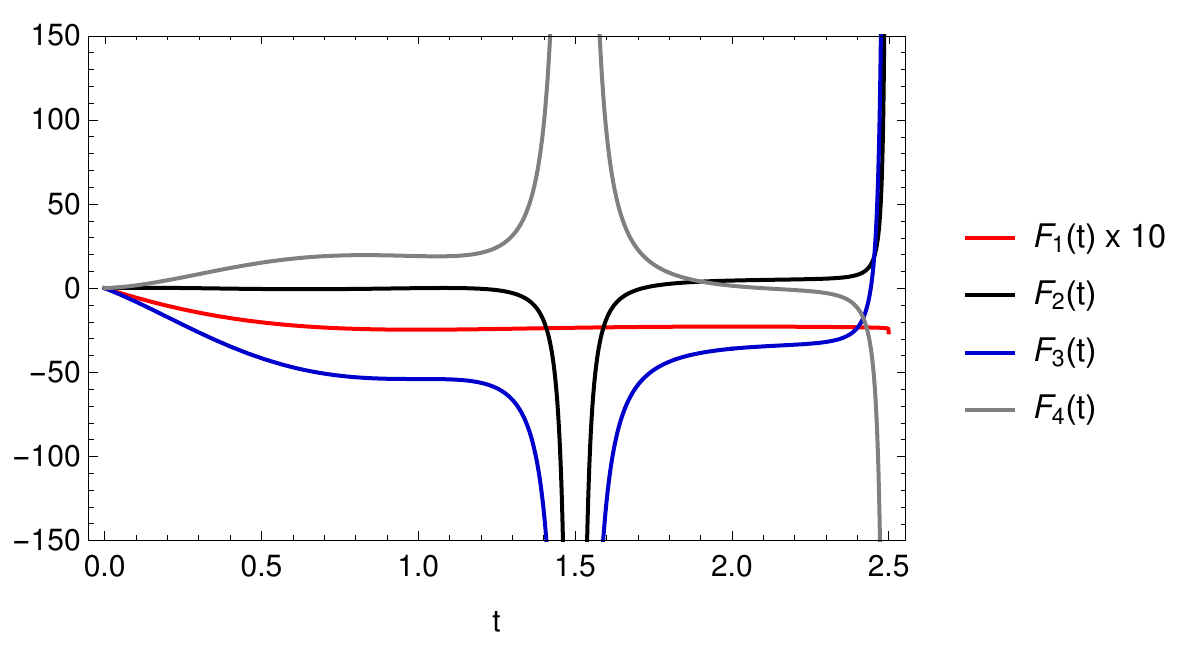}
    \end{center}
    \caption{The functions $F_1$, $F_2$, $F_3$ and $F_4$ given in Eqs~\eqref{eq:F1},~\eqref{eq:F2},~\eqref{eq:F3},~\eqref{eq:F3}, resp.
    $F_1$ has the first (logarithmic) singularity at $t=5/2$, whereas $F_{2-4}$ show a second-order pole 
    at $t=3/2$.
    }
    \label{fig:Fs}
\end{figure}

We show the functions $F_1$, $F_2$, $F_3$ and $F_4$ in Fig.~\ref{fig:Fs}. 
In particular, we notice that functions $F_{2-4}$ feature the first
singularity at $t=3/2$, which is not present in the $\mathcal{O}(1/N)$ critical exponents 
but shows up at the $\mathcal{O}(1/N^2)$ level. This singularity
gets exactly cancelled in the combination of $F_3$ and $F_4$ 
entering $\tilde{\omega}_+^{(1)}$, Eq.~\eqref{eq:omplus}, while
$F_2$ does not contribute to $\omega_{\pm}^{(1)}$.

\subsection{Perturbative results}
We provide here the expansion of the functions $F_{1-4}$ in order to check against
the known perturbative results and predict the $\mathcal{O}(1/N)$
terms that would appear up to six-loop order.

In the following, we give the explicit term-by-term expansions of the $\beta$-functions, given in Eqs~\eqref{eq:beta1} 
and~\eqref{eq:beta2}. First, for the Yukawa $\beta$-function we have
\begin{align}
 y^2 F_1(y N) = & - 6 y^3 N  + \frac{7}{2} y^4 N^2 + \frac{11}{6} y^5 N^3 \nonumber\\ 
 & + \left( \frac{19}{16} - 3 \zeta_3 \right) y^6 N^4 \\ 
 & +\left( \frac{7}{8}   + \frac{14}{5} \zeta_3 - \frac{18}{5} \zeta_4\right) y^7 N^5 +
 \mathcal{O}(y^8).\nonumber 
\end{align}
As for the quartic-coupling $\beta$-function, 
Eq.~\eqref{eq:beta2}, one can predict the coefficients
based on $F_2$, $F_3$ and $F_4$. Their contribution to the $\beta$-function is:
\begin{align}
 y^2 F_2(y N) &= 4 y^3 N - \frac{157}{8} y^4 N^2 + \left( 42 \zeta_3 - \frac{193}{6} \right) y^5 N^3\nonumber \\ 
 & \ + \left(  - \frac{2623}{64} - \frac{157}{4} \zeta_3+90\zeta_4 \right) y^6 N^4 \\ 
 & \ +\left( - \frac{3993}{80} - \frac{491}{5}\zeta_3 -\frac{426}{5}\zeta_4 + 234 \zeta_5 \right)y^7 N^5\nonumber\\
 &\ + \mathcal{O}(y^8),\nonumber
\end{align}
\begin{equation}\begin{split}
 \lambda^2 F_3(y N) = & -72 y \lambda^2 N - 108 y^2 \lambda^2 N^2 \\ &
 + 144( 2 \zeta_3-1) \lambda^2 y^3 N^3 \\ &
 -180( 1 + 2 \zeta_3 -3\zeta_4) \lambda^2 y^4 N^4 \\ &
 -216(1 + 2 \zeta_3 +3\zeta_4- 6 \zeta_5)\lambda^2 y^5 N^5 \\
 & + \mathcal{O}(\lambda^2 y^6),
 \end{split}
\end{equation}
\begin{align}
y \lambda  F_4&(y N) = 7 \lambda y^2 N+\frac{217}{2} \lambda y^3 N^2 \nonumber\\
    &\ + \left( \frac{1685}{12} - 228 \zeta_3 \right) \lambda y^4 N^3 \\ 
    &\  + \left( \frac{699}{4}  + 248 \zeta_3 - 450 \zeta_4\right) \lambda y^5 N^4 \nonumber\\ 
    &\  +\left( \frac{3359}{16}  +
	\frac{2123}{5} \zeta_3 + \frac{2409}{5} \zeta_4- 1116 \zeta_5 \vphantom{\frac12}\right) \lambda y^6 N^5\nonumber\\
    &\ + \mathcal{O}(\lambda y^7).\nonumber
\end{align}

We have checked that the expansions up to $\mathcal{O}(N^3)$ agree with the known leading-$N$ four-loop perturbative 
result~\cite{Zerf:2017zqi}. The $\mathcal{O}(N^4)$ and $\mathcal{O}(N^5)$ terms are the leading-$N$ 
prediction for the five- and six-loop terms, respectively.

\FloatBarrier
\section{Conclusions}
\label{sec:conclusions}
Our goal in this paper was to compare the critical-point and bubble-resummation
methods for computing the $\beta$-functions in the large-$N$ limit. While for the single-coupling 
theories, the methods are equivalent, we have shown that in the multi-coupling case the critical 
exponents are only sensitive to specific combinations of the different functions entering the 
$\beta$-functions, 
and direct computations by means of bubble resummations or additional information are needed to 
decipher this missing information. On the other hand, the critical-point method is more powerful 
for obtaining results beyond the leading $1/N$ order, and thereby the two methods can provide complementary
information.

However, we envisage that it might be possible to reconstruct the full system of RG functions in a multiple-coupling theory within the 
critical-point formalism through extraction of the operator-product-expansion (OPE) coefficients. 
A detailed study of the three-point function Schwinger--Dyson equation would be necessary, from which one could extract the 
OPE consistently at every order in $1/N$ in a similar fashion as for the critical exponents. This is in line with 
the recent analyses carried out in the functional-RG framework~\cite{Codello:2017qek,Codello:2017hhh,Codello:2018nbe,Codello:2019vtg}.

Concretely, we have presently computed the $\beta$-function for the quartic coupling in the GNY model, thereby completing the 
computation of the full system at $\mathcal{O}(1/N)$ level. While the critical exponents $\omega_{\pm}$ computed recently
up to $\mathcal{O}(1/N^2)$ imply that the radius of convergence of the critical exponents shrinks from $\mathcal{O}(1/N)$ to $\mathcal{O}(1/N^2)$, 
our present result shows that this new pole only appearing in $\omega_{\pm}^{(2)}$ is actually already present at 
the $\mathcal{O}(1/N)$ level, when the full system of $\beta$-functions is known. We showed that the disappearance of the pole 
is due to a subtle cancellation between the various resummed function in the computation of the critical exponents. 

For the one-coupling case we have briefly revisited the question about the possible IR fixed point of the two-dimensional GN
model by studying the $\mathcal{O}(1/N^2)$ $\beta$-function. This complements the previous studies using the four-loop perturbative
results with the help of Pad\'{e} approximants. We restate that there is no indication 
of an IR fixed point.

\section*{Acknowledgements}
We thank John Gracey  for valuable discussions and Anders Eller Thomsen for the participation in the initial stage of this project.
The CP$^3$-Origins centre is partially fundedby the Danish National Research Foundation, grant number DNRF:90.
\appendix
\section{Gross--Neveu--Yukawa $\beta$-functions at $1/N^2$}
\label{app:N2}
The $\mathcal{O}(1/N)$ ansatz for the GNY $\beta$-functions in Eqs~\eqref{eq:beta1} and
\eqref{eq:beta2} is extended to $\mathcal{O}(1/N^2)$ as
\begin{equation}\begin{split}
 \beta_y = & - \epsilon y 
 + y^2 [ 2 N + 3 + F_1(y N) - y( 9/8 + F_5(y N)) ] \\ &
 - y^2 \lambda (24 + F_6(y N)) + 
 y \lambda^2(24 + F_7(y N)),
 \end{split}
\end{equation}
\begin{equation}\begin{split}
 \beta_\lambda = & - \epsilon \lambda
 + y^2 [ - N + F_2(y N) + y F_8(y N) + \lambda F_9(y N) ] \\ &
 + \lambda^2 [36 + F_3(y N) -\lambda( 816 + F_{11}(y N)) ] \\ &
 + y \lambda [4 N + F_4(y N) + \lambda F_{10}(y N) ],
 \end{split}
\end{equation}
where we have introduced seven new unknown
functions, $F_{5-11}$, such that $F_{5-11}(0) =0$.
We then compute the Jacobian for $(\beta_y,\beta_\lambda)$
and we evaluate it at the Wilson--Fisher fixed point. The eigenvalues are then
matched with the critical exponents $\omega_\pm$.
At $\mathcal{O}(1/N^2)$, we find the following relations:
\begin{equation}
\begin{split}
 & -57
 - 4 \left[ 3 + F_1 \left(\epsilon\right)
 \right] \left[ 2 F_1^\prime \left(\epsilon\right) 
 +  \epsilon F_1^{\prime\prime} \left(\epsilon\right) \right]
 \\ &  - 
 8 F_5 \left(\epsilon\right)
 - 2 F_7 \left(\epsilon\right) -8 \epsilon
 F_5^\prime \left(\epsilon\right) \\ &
 -4 \epsilon F_6^\prime \left(\epsilon\right)
 + 2 \epsilon F_7^\prime \left(\epsilon\right) =
8 \frac{ \tilde{\omega}_-^{(2)}(2 \epsilon)}{\epsilon^2}.
 \end{split}
\end{equation}
and 
\begin{align}
& -2520 - 4782 \epsilon
+ 8 \epsilon F_{10}(\epsilon)
-6 \epsilon F_{11}(\epsilon)
- 288 F_2(\epsilon) \nonumber \\ 
&- 144 F_3(\epsilon)
-8 F_2(\epsilon)F_3(\epsilon)
- 2 F_3(\epsilon)^2
-156 F_4(\epsilon) \nonumber\\ 
&- 4 F_3(\epsilon) F_4(\epsilon)
+ 16 \epsilon F_5(\epsilon) 
+ 8 \epsilon F_6(\epsilon)
-4 \epsilon F_7(\epsilon) \nonumber\\
&+ 8 \epsilon F_9(\epsilon) 
- 12 \epsilon F_3^\prime(\epsilon)
- 12 \epsilon F_4^\prime(\epsilon)
+ 48 F_1(\epsilon) \\
&+ 24 \epsilon F_1^\prime(\epsilon)
+ 8 F_1(\epsilon)^2 
- 4 F_1(\epsilon) F_4(\epsilon) +
8 \epsilon F_1(\epsilon) F_1^\prime(\epsilon)\nonumber \\
&- 4 \epsilon F_1(\epsilon) F_3^\prime(\epsilon)
- 4 \epsilon F_1(\epsilon) F_4^\prime(\epsilon)
= 8 \frac{\tilde{\omega}_+^{(2)}\left(2 \epsilon\right)}{\epsilon}.\nonumber
\end{align}
Clearly, one needs more input than $\omega_\mp^{(2)}$
to obtain $F_{5-11}$. On the other hand,
direct computation at $\mathcal{O}(1/N^2)$
is technically challenging and beyond the scope
of this paper.

\bibliography{refs.bib}
\onecolumngrid
\twocolumngrid

\end{document}